\documentclass[preprint, 12pt]{elsarticle}

\usepackage{amssymb}
\usepackage{amsthm}
\usepackage{amsmath}
\usepackage{lineno}
\usepackage{xspace}
\usepackage{color}
\def\uPIC {$\mu$-PIC\xspace}

\begin{document}
\begin{frontmatter}

\title{Triaxial \uPIC for resolving Multi-Hit Ambiguity}

\author[Kyoto]{A.~Takada\corref{cor}}
\cortext[cor]{Corresponding author}
\ead{takada@cr.scphys.kyoto-u.ac.jp}

\author[Kyoto]{M.~Abe}
\author[Kyoto]{Y.~Yoshida}
\author[Kyoto]{T.~Ikeda}
\author[Cross]{J.~D.~Parker}
\author[Kobe]{K.~Miuchi}
\author[DNP]{K.~Ohta}
\author[DNP]{T.~Motomura}
\author[Kyoto]{T.~Tanimori}

\affiliation[Kyoto]{
    organization = {Graduate School of Science, Kyoto University},
    addressline  = {Kitashirakawa-Oiwakecho},
    city         = {Kyoto},
    postcode     = {606-8502},
    state        = {Kyoto},
    country      = {Japan}
}
\affiliation[Cross]{
    organization = {Neutron Science and Technology Center, CROSS},
    addressline  = {162 Shirakata},
    city         = {Tokai},
    postcode     = {319-1106},
    state        = {Ibaraki},
    country      = {Japan}
}
\affiliation[Kobe]{
    organization = {Graduate School of Science, Kobe University},
    addressline  = {1-1 Rokkodai, Nada},
    city         = {Kobe},
    postcode     = {657-8501},
    state        = {Hyogo},
    country      = {Japan}
}
\affiliation[DNP]{
    organization = {Dai Nippon Printing Co., Ltd.},
    addressline  = {250-1 Wakashiba},
    city         = {Kashiwa},
    postcode     = {277-0871},
    state        = {Chiba},
    country      = {Japan}
}

\begin{abstract}
The micro pixel chamber (\uPIC), a type of micro-pattern gaseous detectors (MPGD), 
features pixel electrodes structured similarly to a sliced proportional counter. 
Conventionally, \uPIC electrodes are connected to orthogonal strip electrodes 
to reduce the number of readout circuits; 
however, this configuration introduces multi-hit ambiguity 
when tracing winding tracks or during high-rate operations.
A well-known solution to this problem is the adoption of an additional readout axis. 
To this end, we designed a new electrode structure using Garfield++ simulations 
and fabricated a triaxial \uPIC based on through-glass-via (TGV) technology. 
This triaxial \uPIC achieved a high gas gain of $1.1\times 10^4$
and the energy resolution of 22\% (FWHM) at 5.9~keV,
Furthermore, the false-hit probability was less than 0.78\% at a 95\% confidence level
for two clusters separated by a distance of over 3 mm, 
consistent with Garfield++ predictions. 
With this triaxial \uPIC, 
single avalanche signals are collected 
from three independent readout strip electrodes (the anode and two cathodes). 
The charge ratio among these electrodes remains nearly constant at anode:U-cathode:V-cathode=3:2:1.
In addition, we successfully achieved an enhanced gain of $3.2\times 10^4$ 
without increasing the discharge risk by employing an alternative operation mode.
\end{abstract}


\end{frontmatter}


\section{Introduction}
A micro pixel chamber (\uPIC)~\cite{2001NIMPA.471..264O}, 
the development of which began in the early 2000s, 
is a gaseous 2D imaging detector featuring fine pixel electrodes 
similar to a sliced proportional counter. 
With an anode diameter of approximately 60~$\mu$m and an anode-cathode gap of 100~$\mu$m,
it achieves a high stable gain exceeding 6,000~\cite{2004NIMPA.535..236M}.
Since the electrodes are fabricated using printed circuit board (PCB) technology, 
the \uPIC offers a fine spatial resolution of 120~$\mu$m (RMS)~\cite{2004NIMPA.525...20N}
and a large detection area of 30$\times$30~cm$^2$~\cite{2007NIMPA.573..195T}.
{\uPIC}s are currently employed in various applications, 
including MeV gamma-ray astronomy~\cite{2004NewAR..48..263T, 2022ApJ...930....6T}, 
neutron imaging~\cite{2004NIMPA.529..373T},
dark matter search~\cite{2007PhLB..654...58M}, medical imaging~\cite{2010NIMPA.623..606K},
space dosimetry~\cite{2013NIMPA.732..591K}, 
active targets for radioisotope beam experiments~\cite{2014JPhCS.569a2042F},
and environmental monitoring~\cite{2017NatSR...741972T}.
Recently, some applications have demanded a higher stable gain of over 15,000. 
Previous simulation studies suggest that a substrate thicker than 200~$\mu$m 
can provide a gain several times higher by suppressing the electric field produced 
by the backside anode strips~\cite{2005NIMPA.546..457N}.
However, it is challenging to fabricate narrow anode vias 
through substrates exceeding 100~$\mu$m in thickness. 
Consequently, we have begun developing {\uPIC}s based on 
micro-electro-mechanical systems (MEMS) technology~\cite{2018EPJWC.17402010T, 2020JPhCS1498a2002A}, 
which enables the production of more accurate electrodes on thicker substrates compared to PCB technology. 
In particular, a through-glass-via (TGV) \uPIC achieves twice the gain of a PCB \uPIC 
at the same anode voltage, 
while demonstrating high resilience against discharge-induced damage or failure~\cite{2020JPhCS1498a2002A}.

\begin{figure}
    \centering
    \includegraphics[width=\linewidth]{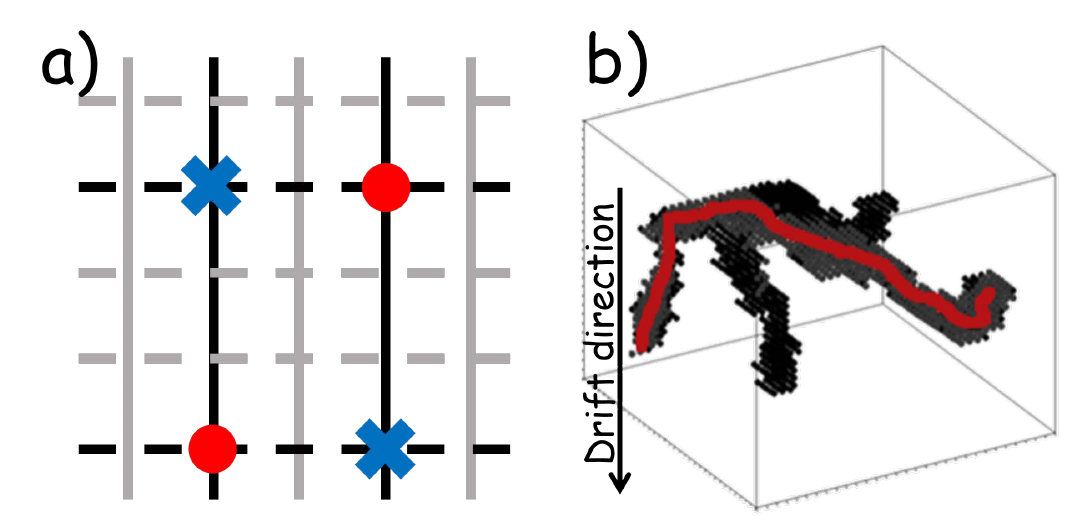}
    \caption{
        Orthogonal strip readout electrodes and multi-hit ambiguity.
        a) When two signals (red-filled circles) come into the detector at the same time, 
        four candidates exist by taking the coincidence of the black hit-strips. 
        We cannot distinguish between two true-hit points and two fake-hit points (blue crosses).
        b) In the case of a winding track of charged particles like the red curve, 
        a time projection chamber based on the orthogonal strip readouts 
        provides the black track by taking the coincidence of two projected images.
        We cannot distinguish between the true track and the fake track.
    }
    \label{fig:orthogonal_strip}
\end{figure}

\begin{figure}
    \centering
    \includegraphics[width=\linewidth]{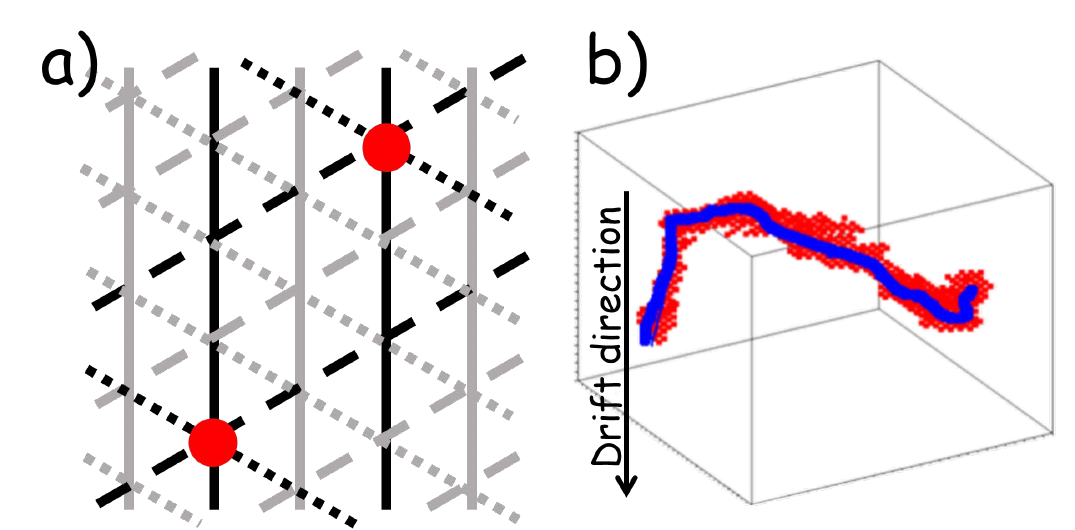}
    \caption{
        Hexagonal strip readout electrodes for solving multi-hit ambiguity.
        a) Even if two signals (red-filled circles) come into the detector at the same time, 
        we obtain the correct pit-pixel position 
        by taking a coincidence of three hit-strip electrodes (black line).
        b) Using these hexagonal strip readout electrodes, 
        we can observe a winding track (blue line) of a charged particle without any fake tracks.
    }
    \label{fig:hexagonal_strip}
\end{figure}

On the other hand, 
large-area micro-pattern gas detectors require a large number of readout circuits. 
For instance, a $30\times30$~cm$^2$ \uPIC contains $5.9\times10^5$ proportional counters.
To reduce the circuit count, each pixel electrode of the \uPIC is 
connected to orthogonal readout strips instead of individual pixel readouts; 
thus, a $30\times30$~cm$^2$ \uPIC consists of 
768 anode strips (X-axis) and 768 cathode strips (Y-axis). 
While the adoption of strip readout facilitates the use of large {\uPIC}s in various applications, 
it also leads to multi-hit ambiguity. When two signals enter the detector simultaneously,
four candidate hits are generated by taking the coincidence of the hit strips, 
as shown in Figure~\ref{fig:orthogonal_strip}a.
It is then impossible to distinguish between the two true-hit points (filled circles) and the two fake-hit points (ghost hits, indicated by crosses).
This multi-hit ambiguity becomes a serious issue 
when observing winding electron tracks (Figure~\ref{fig:orthogonal_strip}b) or during high-rate operation. 
A well-known solution to this problem is the adoption of an additional readout axis. 
As shown in Figure~\ref{fig:hexagonal_strip}a, 
even if two signals (filled circles) enter the detector simultaneously, 
the correct pixel positions can be identified by taking a coincidence of three hit-strip electrodes. 
Consequently, winding tracks of charged particles can be observed 
without fake tracks (Figure~\ref{fig:hexagonal_strip}b). 
Many cases of multi-hit ambiguity have been resolved using this approach, 
and some detectors~\cite{2002NIMPA.478..104B, 2005NIMPA.553...18S}
already incorporate an additional axis electrode. 
However, the readout pads, wires, or electrodes in these detectors are arranged so as not to overlap, 
picking up signals induced by different electrons at different positions. 
As a result, the signal ratio for each electrode remains unstable 
due to fluctuations in charge division caused by electron diffusion.

Monte Carlo simulations are instrumental in studying avalanches 
and induced signals in micro-pattern gas detectors. 
In the early stages of \uPIC development, 
Garfield~\cite{1998NIMPA.419..726V}-—a macroscopic-scale detector simulator-—
could describe the gas gain ratios under different anode voltages
but failed to simulate absolute gas gain values~\cite{2005NIMPA.546..457N, 2005NIMPA.540..266B}.
Recently, Garfield++~\cite{Garfield_pp} was developed for microscopic-scale simulations 
and can now accurately describe the absolute avalanche size in PCB {\uPIC}s~\cite{2013JInst...8C0023T}.
For detectors with insulating substrates, 
Garfield++ simulations show excellent agreement with experimental data across various {\uPIC}s,
regardless of substrate material or pixel structure~\cite{2018EPJWC.17402010T, 2020JPhCS1498a2002A}.
This suggests that simulation-based pixel electrode design has become a pragmatic approach, 
replacing conventional design methods that relied heavily on empirical experience.

\begin{figure}
    \centering
    \includegraphics[width=\linewidth]{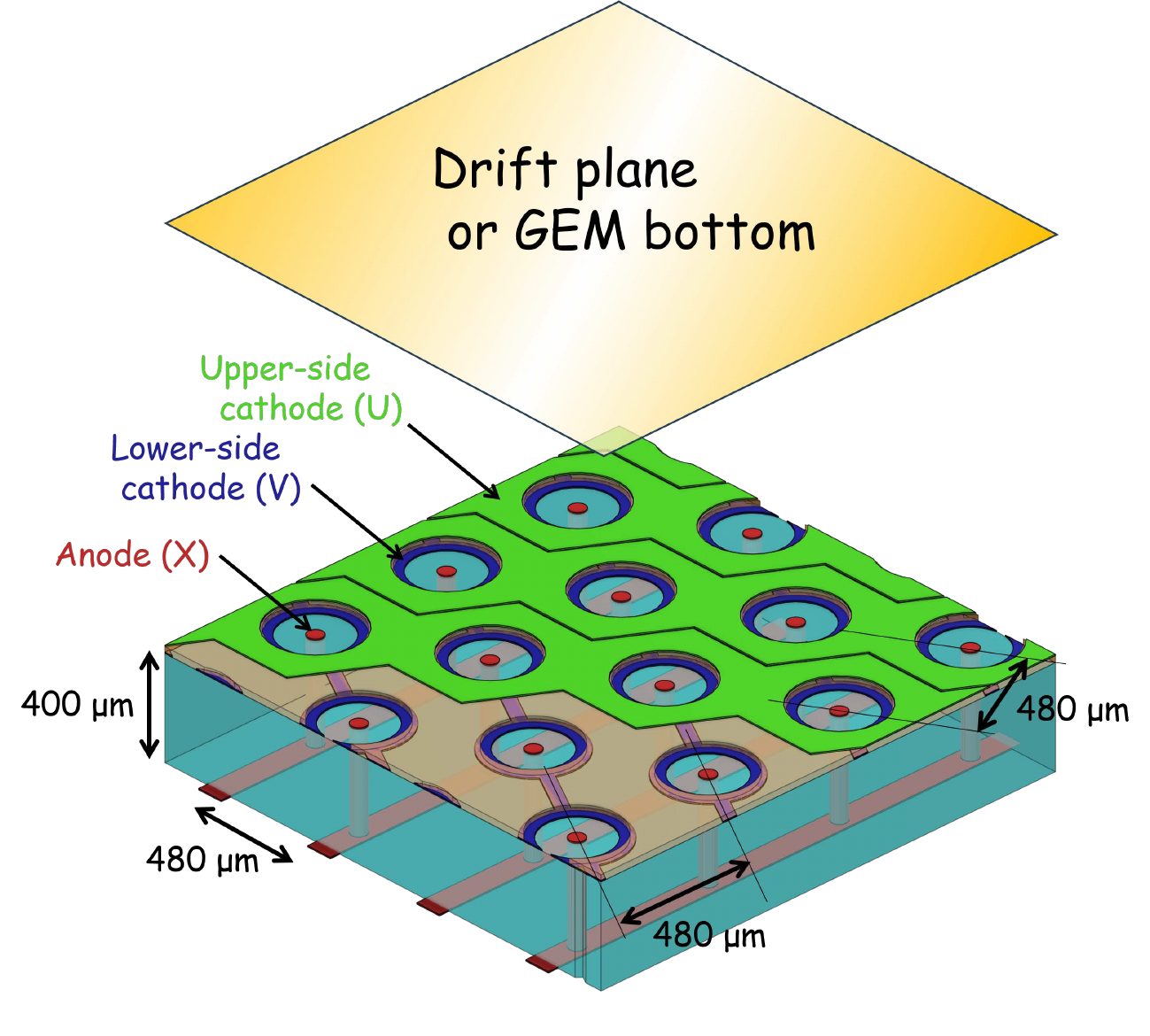}
    \caption{
        Schematic of a triaxial \uPIC. 
        A gas avalanche is generated near the anode, 
        and the produced ion-electron pairs induce charge currents 
        in the anode, lower cathode, and upper cathode of the same pixel, respectively.
    }
    \label{fig:geometry}
\end{figure}
To resolve the multi-hit ambiguity, 
we designed a new electrode structure featuring three readout strips 
with different orientations, as shown in Figure~\ref{fig:geometry} (Section~\ref{sec:design}).
We also fabricated this new \uPIC (triaxial \uPIC) using the TGV process 
and evaluated its avalanche size and charge division ratio (Section~\ref{sec:norm_op}).
Furthermore, we verified the extent to which the triaxial \uPIC resolves multi-hit ambiguity. 
Finally, we demonstrate an alternative operation mode to achieve higher gain (Section~\ref{sec:alt_op}).

\begin{figure}
    \centering
    \includegraphics[width=\linewidth]{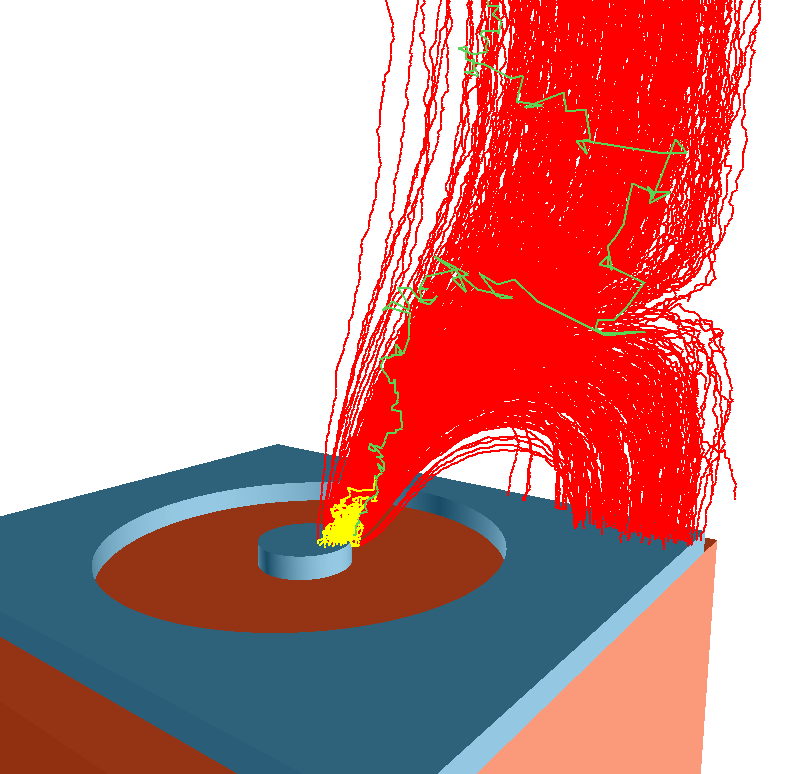}
    \caption{
        Typical avalanche for single seed electron simulated by Garfield++.
        The pixel electrode structure is assumed to be the PCB \uPIC.
        The seed electron is represented by the green line.
        Yellow and red lines are the tracks of electrons and ions produced by avalanches, respectively.
        The avalanche grows in the specific direction in which the seed electron approached the anode electrode.
    }
    \label{fig:avalanche_single}
\end{figure}

\section{Design of a triaxial \uPIC}
\label{sec:design}
\begin{figure}
    \centering
    \includegraphics[width=\linewidth]{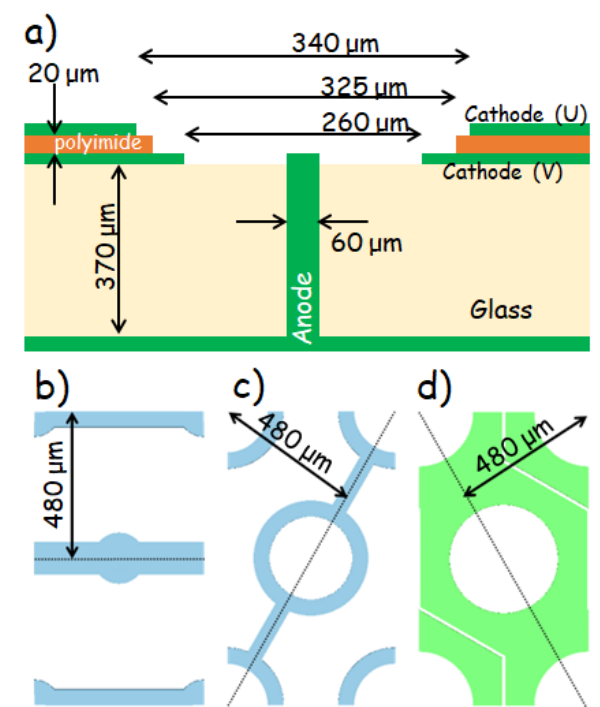}
    \caption{
        Schematic views of the basic electrode structure in triaxial \uPIC. 
        a) Cross-section image of pixel electrodes; 
        b) anode strips placed at the backside of the glass substrate; 
        c) lower-side cathode (V-cathode) strips placed on the glass substrate; 
        d) upper-side cathode (U-cathode) strips placed on the polyimide. 
        The dotted line in b), c), and d) represents the center of each strip.
    }
    \label{fig:sim_structure}
\end{figure}

There are two primary methods for incorporating a third readout axis.
The first is employing an insulated back-strip electrode 
to pick up the induced charge~\cite{1996NIMPA.381..280T}. 
The second is using a divided electrode structure. 
Although the back-strip method is easier to fabricate as it requires no major structural modifications,
signal detection is challenging because the charge induced on the back-strip is 
only approximately one-tenth of that on the other electrodes. 
In contrast, while the divided electrode splits the induced charge, 
the signal remains large as it is induced directly by the electron-ion motion. 
Furthermore, the charge division ratio can be precisely controlled by the electrode design. 
Therefore, we adopted the divided electrode method to ensure compatibility 
with our existing data-acquisition system~\cite{2015NIMPA.800...40M}.
At a \uPIC pixel, 
the avalanche grows in the specific direction from which the seed electron approaches the anode, 
as illustrated in Figure~\ref{fig:avalanche_single}.
Thus, an axially symmetric electrode structure is desirable to maintain a constant charge division fraction.
To achieve this, we divided the cathode into concentric electrodes. 
We adopted a vertical arrangement for these concentric electrodes 
to prevent a decrease in collection efficiency (Figure~\ref{fig:sim_structure}a), 
which would otherwise occur if seed electrons fell into the gaps between the cathode electrodes.
To minimize the risk of malfunction, 
the designs of the anode and the lower cathode follow the electrode structure 
of the TGV \uPIC without polyimide coating,
where the diameters of the anode and the V-cathode inner-hole are 60 and 260 $\mu$m, respectively. 
The pixels are arranged in a close-packed structure to capture detailed tracks of charged particles, 
with the differential angle between the strip electrodes of each axis set to 120$^\circ$.

\begin{figure}
    \centering
    \includegraphics[width=\linewidth]{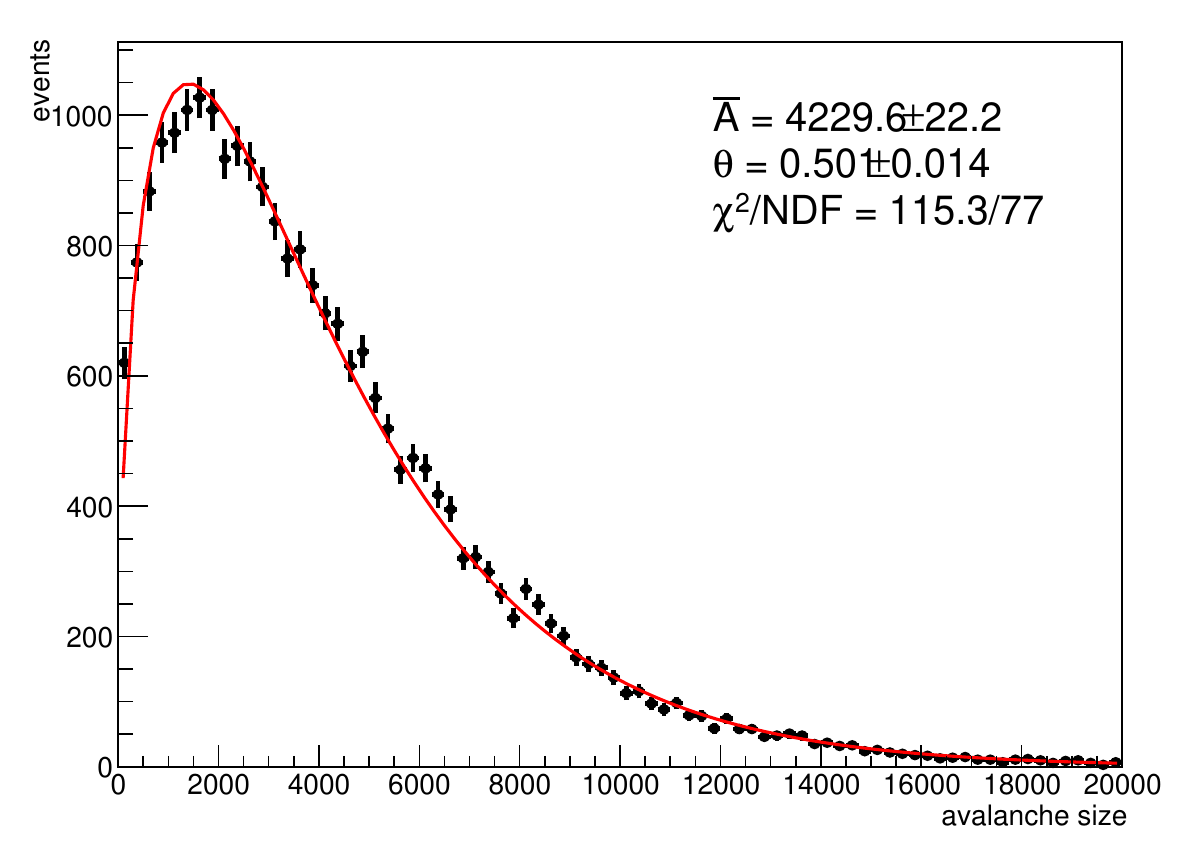}
    \caption{
        Single-electron spectrum of a triaxial \uPIC 
        with an anode voltage of 490~V and U/V cathode voltage of 0~V.
        Assumed gas is Ar 90\% + C$_2$H$_6$ 10\% at a pressure of 1~atm.
        The filled circles are obtained by Garfield++ simulation, 
        and the solid line is the best fit with a Polya distribution.
    }
    \label{fig:single_electron}
\end{figure}

\begin{figure}
    \centering
    \includegraphics[width=\linewidth]{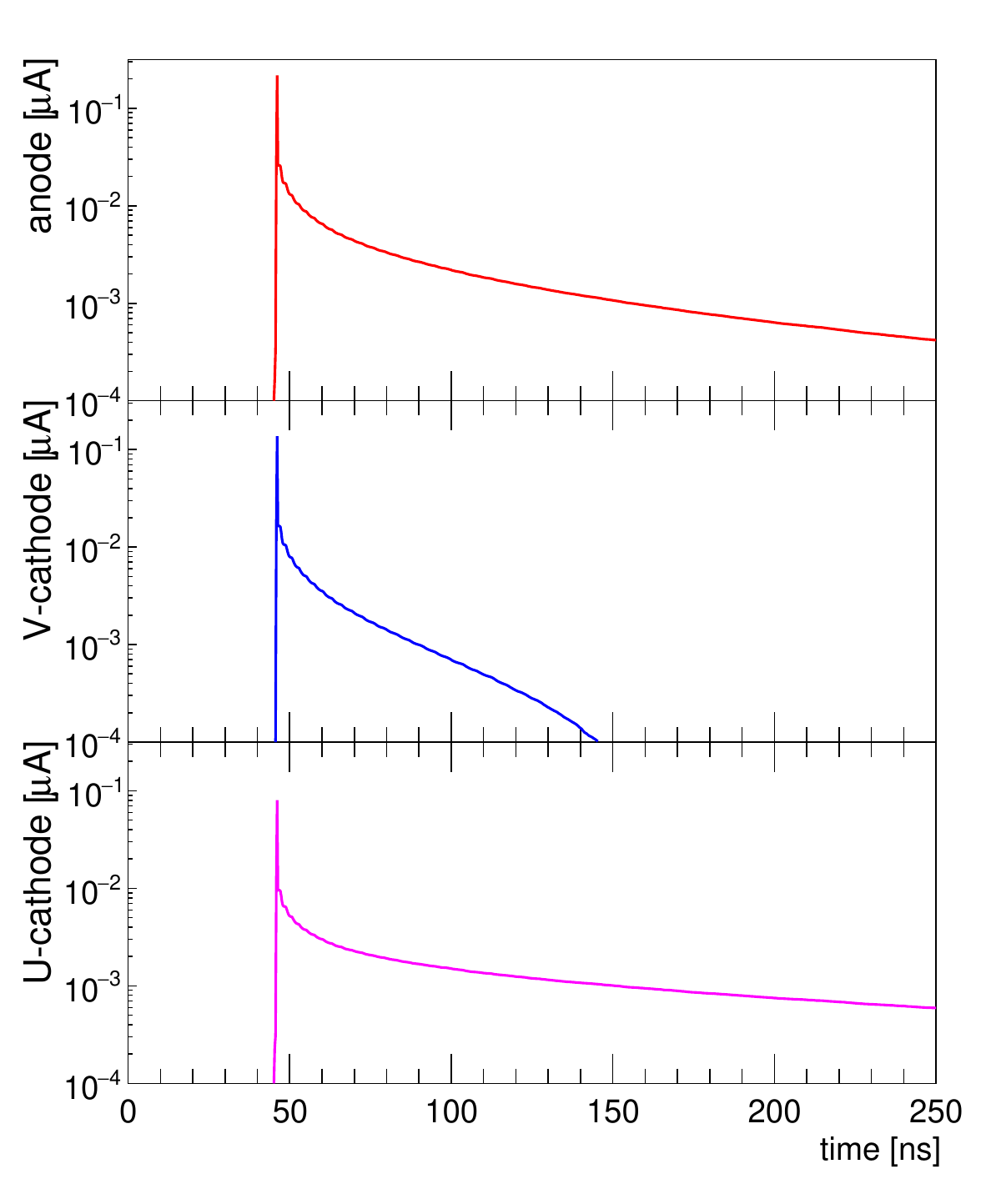}
    \caption{
        Typical induced current (absolute values) in each electrode for a single seed electron at a gas gain of 4000, simulated using Garfield++. 
        All waveforms have a sharp spike induced by electrons and a slow tail induced by ions. 
        The tail of the V-cathode signal waveform is shorter than the tails of other electrode signals 
        because the signals induced by ions are limited to ions moving 
        through the neighborhood of the narrow bare electrode of the V-cathode.
        The U-cathode signal has a long tail induced by ions flowing back into the drift region.
    }
    \label{fig:induced_current}
\end{figure}

\begin{figure}
    \centering
    \includegraphics[width=\linewidth]{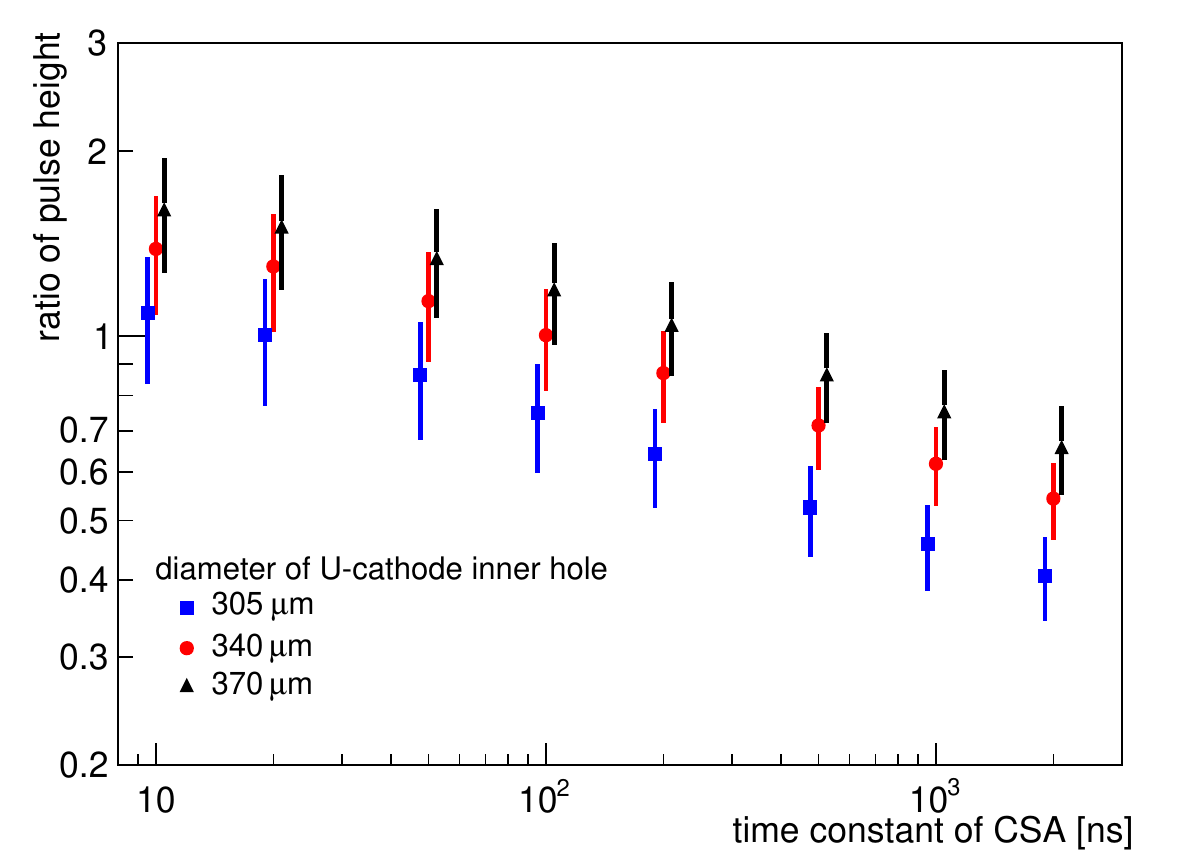}
    \caption{
        Ratio of the pulse height of the V-cathode for a single seed electron to that of the U-cathode 
        as a function of a time constant of simple CSA with the different diameters of the U-cathode inner hole,
        obtained from Garfield++ simulation.
        For visibility, the time constants of the square and triangle are set to -5\% and +5\%, respectively.
    }
    \label{fig:sim_v_cathode}
\end{figure}

To verify the basic design of the triaxial \uPIC, 
we simulated the avalanche process and the induced current on each electrode using Garfield++ (with Magboltz v11).
For the calculations of the 3D electric and weighting fields, 
we utilized ElmerSolver (v9.0)~\cite{Elmer_HP} based on 
a finite element mesh generated by Gmsh (v4.11.1)~\cite{Gmsh_HP}. 
The Penning transfer probability in this study was fixed at 0.35. 
At this value, the simulated avalanche size matches the measured gas gain 
at anode, U-cathode, and V-cathode voltages of 440 V, 0 V, and 0 V, respectively.
Figure~\ref{fig:single_electron} shows the simulated single-electron spectrum, 
assuming an anode voltage of 490~V, both cathode voltages of 0~V, 
and a gas mixture of Ar (90\%) + C$_2$H$_6$ (10\%) at 1~atm. 
As with previous {\uPIC}s~\cite{2013JInst...8C0023T}, 
the single-electron spectrum follows a Polya distribution. 
The typical induced current waveforms for each electrode are shown in Figure~\ref{fig:induced_current}. 
All waveforms exhibit a sharp spike induced by electrons followed by a slow tail induced by ions, 
consistent with signals observed in PCB {\uPIC}s.
The tail of the V-cathode signal is shorter than those of the other electrodes 
because its ion-induced signal is limited to ions moving 
in the immediate vicinity of the narrow, exposed V-cathode electrode. 
In contrast, the U-cathode signal features a long tail caused by ions drifting back into the drift region.
This difference in the tail components is reflected in the pulse height disparity 
between the U- and V-cathodes; 
furthermore, their ratio depends on the integration time of the readout amplifiers. 
Assuming the use of simple charge-sensitive amplifiers (CSAs), 
Figure~\ref{fig:sim_v_cathode} shows the pulse height ratio of the V-cathode to the U-cathode 
for a single seed electron as a function of the CSA time constant. 
The V-cathode pulse height is nearly equal to that of the U-cathode at a time constant of $\sim$100 ns, 
whereas the U-cathode signal becomes more dominant at larger time constants.
This pulse height ratio also depends on the width of the V-cathode electrode exposed on the surface.
The filled squares, circles, and triangles in Figure~\ref{fig:sim_v_cathode} represent
the pulse height ratios for U-cathode inner-hole diameters of 305, 340, and 370~$\mu$m,
corresponding to exposed V-cathode widths of 15, 32.5, and 47.5~$\mu$m, respectively. 
As the exposed area of the V-cathode becomes narrower, its induced current decreases. 
Based on these results, we concluded that the standard pixel design 
with a U-cathode inner-hole diameter of 340 $\mu$m 
(as shown in Figure~\ref{fig:sim_structure}) is optimal 
for balancing the charge sharing between the V- and U-cathodes.

\begin{figure}
    \centering
    \includegraphics[width=\linewidth]{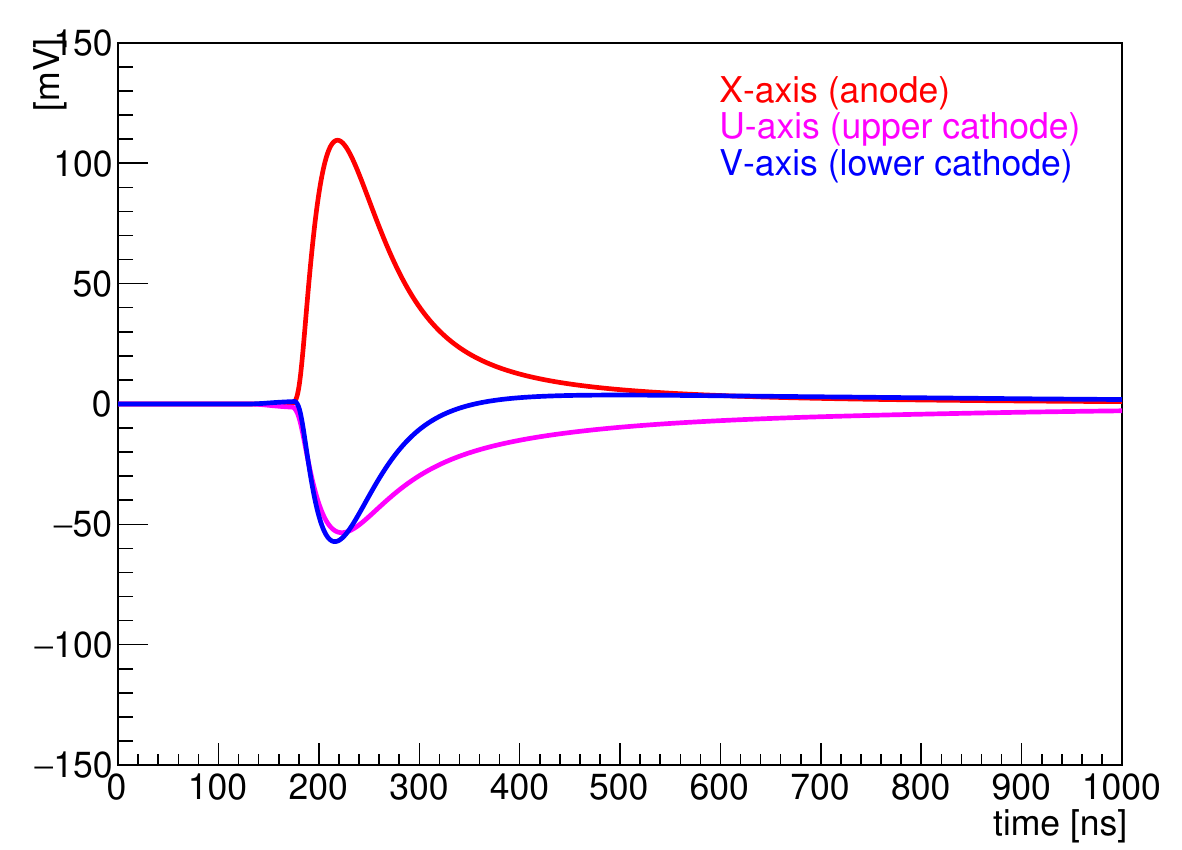}
    \caption{
        Emulated waveform of each electrode under the irradiated X-ray of $^{55}$Fe. 
        The assumed total gas gain is $3.6\times10^4$ 
        in the filled gas of Ar 90\% + C$_2$H$_6$ 10\% at the pressure of 1~atm. 
        The readout circuit is assumed the ASIC and the readout board described in \cite{2015NIMPA.800...40M}.
    }
    \label{fig:emu_wave}
\end{figure}

Finally, we emulated the signal waveforms readout from each electrode.
Figure~\ref{fig:emu_wave} shows the emulated waveforms assuming readout by the ASIC 
and the readout board described in \cite{2015NIMPA.800...40M} 
under irradiation by $^{55}$Fe X-rays, with a total gas gain of $3.6\times10^4$
in an Ar(90\%) + C$_2$H$_6$(10\%) mixture at 1~atm. 
The pulse height of the V-cathode is nearly equal to that of the U-cathode 
and is approximately half that of the anode signal. 
Therefore, a significant increase in amplifier gain is not required to read out both cathodes. 
This compatibility allows for a low-cost upgrade from previous {\uPIC}s
to the triaxial version by reusing the existing operating system.
In some Time Projection Chamber (TPC) applications, 
the energy spectrum is obtained by integrating the signal waveforms. 
Since the pulse width of the V-cathode signal is shorter than those of the U-cathode and anode, 
its integrated charge is expected to be smaller. 
With a sufficiently long integration time, 
the induced charge from ions drifting back into the drift region becomes significant, 
causing the integrated charge of the V-cathode to drop to about one-tenth of that of the other electrodes.
However, in practical operation, 
the integration period is typically limited to the duration 
above a certain threshold to reduce electronic noise; thus, the long tail remains largely unintegrated. 
In the case shown in Figure~\ref{fig:emu_wave}, 
the ratio of the integrated charge readings from the electrodes is approximately 
X:U:V = 2.7:1.8:1 with a threshold of $\pm$10~mV.

\section{Performance}
\label{sec:norm_op}
\subsection{Fundamental test}
\begin{figure}
    \centering
    \includegraphics[width=\linewidth]{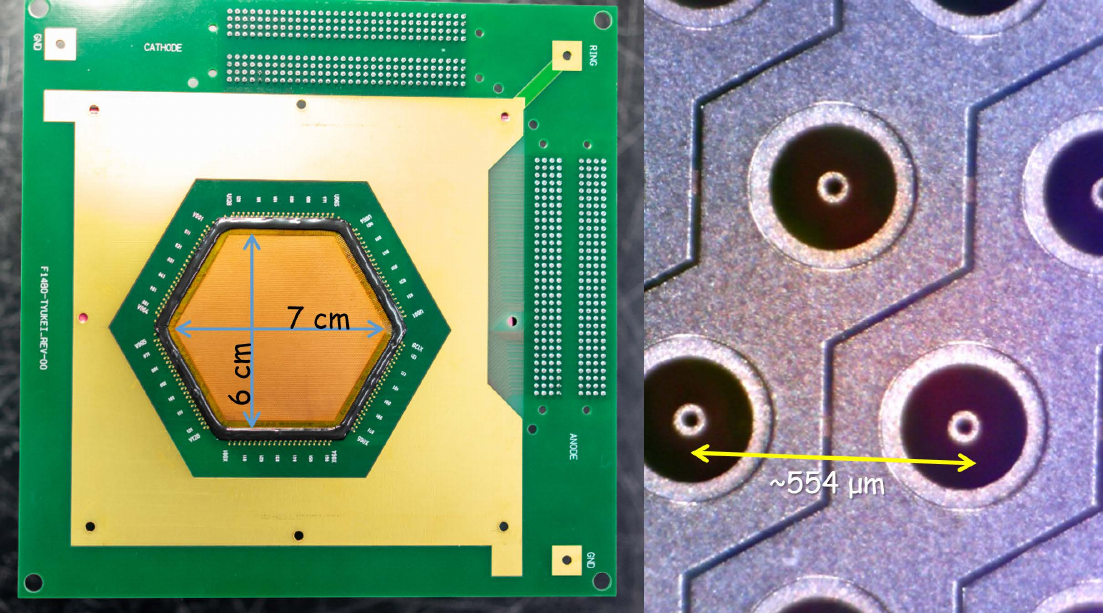}
    \caption{
        (Left) Photo of the fabricated triaxial \uPIC. 
        Each axis consists of 128 strip electrodes, 
        and the overall detection area is hexagonal with a side length of 35.5~mm.
        The strip electrodes of each axis are connected to the readout board 
        via wire bonding at the edges of the hexagon. 
        The coordinates for the X, U, and V axes are defined 
        by the strip IDs of the anode ($i$), U-cathode ($j$), and V-cathode ($k$), respectively.
        These axes are arranged in a counter-clockwise, axially symmetric configuration 
        with a relative angular interval of 120$^\circ$.
        (Right) Micrograph of the pixel electrode of the triaxial \uPIC. 
        The distance between adjacent pixels is $\sim$554~$\mu$m.
    }
    \label{fig:pixel_str}
\end{figure}

\begin{figure}
    \centering
    \includegraphics[width=\linewidth]{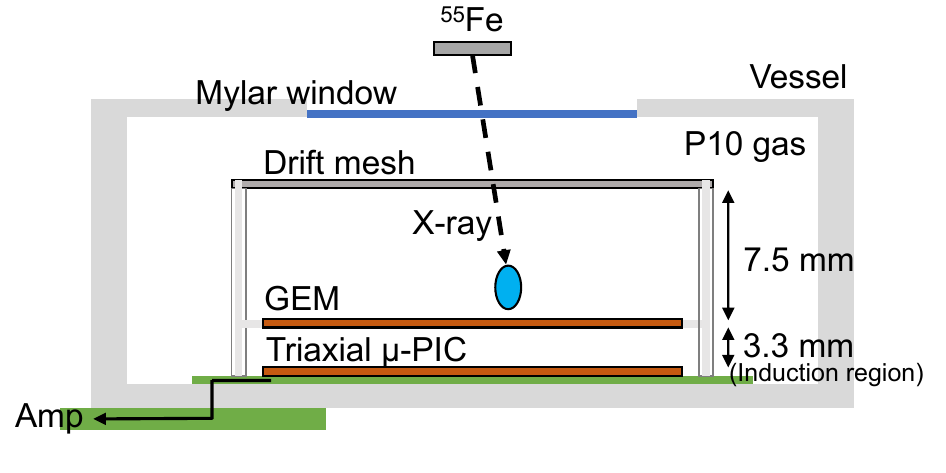}
    \caption{
        Schematic structure of the experimental setup. 
        The triaxial \uPIC was placed at the bottom of the vessel 
        and a GEM was placed above the triaxial \uPIC with an induction gap of 3.3~mm.
        The X-rays were directed into the active volume through a Mylar window. 
        The filled gas was Ar 90\% + C$_2$H$_6$ 10\% at a pressure of 1~atm.
    }
    \label{fig:schematic}
\end{figure}

We fabricated a triaxial \uPIC based on the pixel electrode design described in the previous section. 
Figure~\ref{fig:pixel_str} (left) shows a photograph of the fabricated triaxial \uPIC.
Each axis consists of 128 strip electrodes,
and the overall detection area is hexagonal with a side length of 35.5~mm. 
A micrograph of the pixel electrode is shown in Figure~\ref{fig:pixel_str} (right). 
The electrode structure is identical to the design shown in Figure~\ref{fig:sim_structure},
with a distance between adjacent pixels of approximately 554~$\mu$m. 
A schematic of the experimental setup is shown in Figure~\ref{fig:schematic}.
The triaxial \uPIC was installed at the bottom of a vessel, 
with an $11\times11$~cm$^{2}$ Gas Electron Multiplier 
(GEM~\cite{1997NIMPA.386..531S, 2009NIMPA.608..390T}; SciEnergy Co., Ltd.) placed 3.3~mm 
above it to serve as the first-stage electron amplifier. 
The GEM insulator consists of a 100-$\mu$m-thick liquid crystal polymer, 
with a hole diameter and pitch of 70~$\mu$m and 140~$\mu$m, respectively. 
A mesh electrode was positioned at the top of the active volume 
to create a drift electric field for directing the electron cloud 
into the avalanche section (comprising the GEM and the triaxial \uPIC).
The active region has a thickness of 7.5~mm.
The vessel was filled with an argon-based gas mixture (Ar 90\% + C$_2$H$_6$ 10\%) at a pressure of 1~atm.
The electric fields in the induction region and the active volume were set to 1~kV/cm and 250~V/cm,
respectively, while the voltage difference across the GEM was 320~V.
Under these conditions, the gas gain of the GEM was approximately 9,
which represents a safe operation with no concern for discharge-induced damage.
To read out the signals from the triaxial \uPIC, 
we employed the readout system described in \cite{2015NIMPA.800...40M}.
This readout board encodes the 128-channel \uPIC signals into 4-channel summed waveform data, 
digitized at 50~MHz, and generates a monochromatic track image.

\begin{figure}
    \centering
    \includegraphics[width=\linewidth]{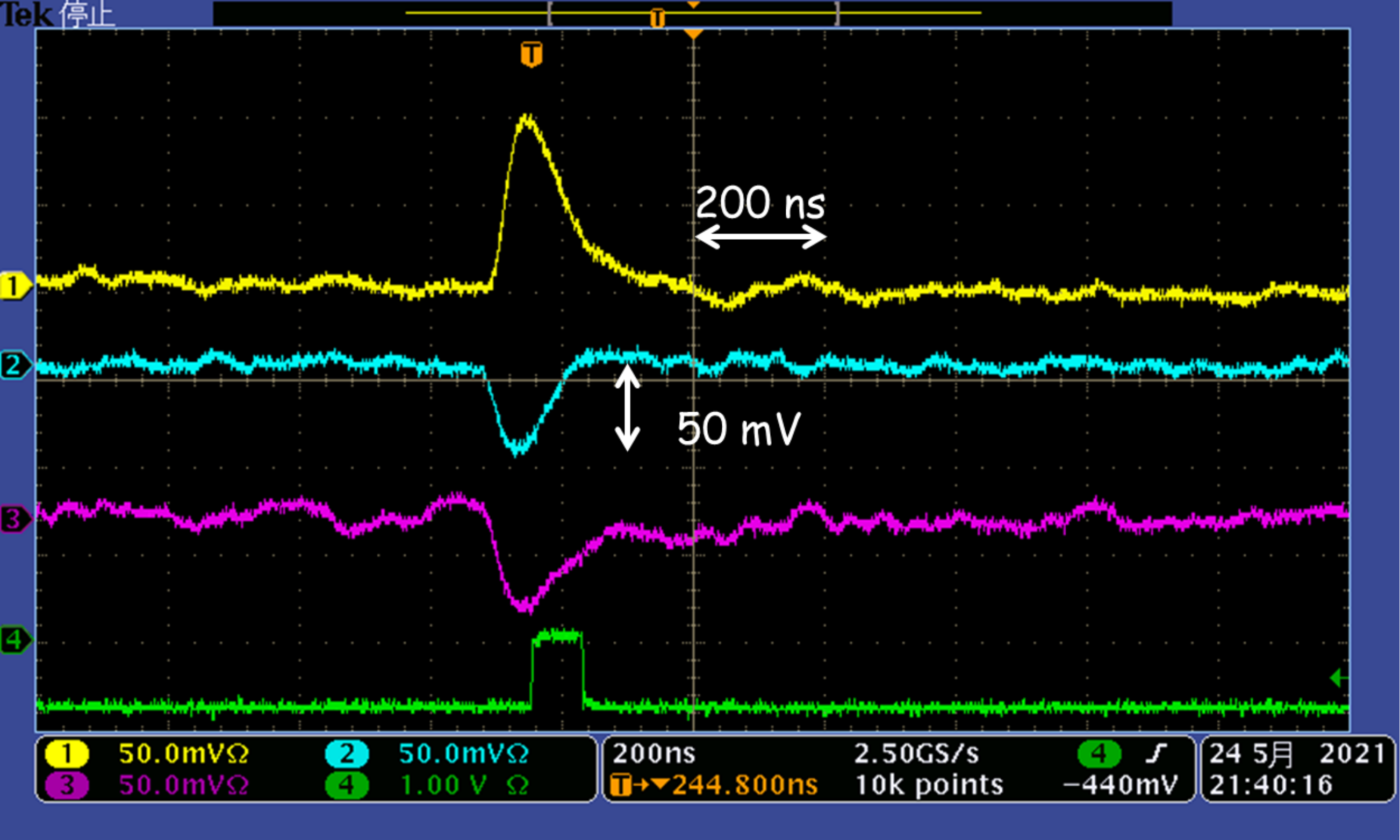}
    \caption{
        Oscilloscope image for the triaxial \uPIC signal at the anode of 490~V using $^{55}$Fe. 
        The yellow, blue and magenta are the anode, V-cathode and U-cathode signals, respectively. 
        The green logic signal indicates the time coincidence.
    }
    \label{fig:waveform}
\end{figure}

\begin{figure}
    \centering
    \includegraphics[width=\linewidth]{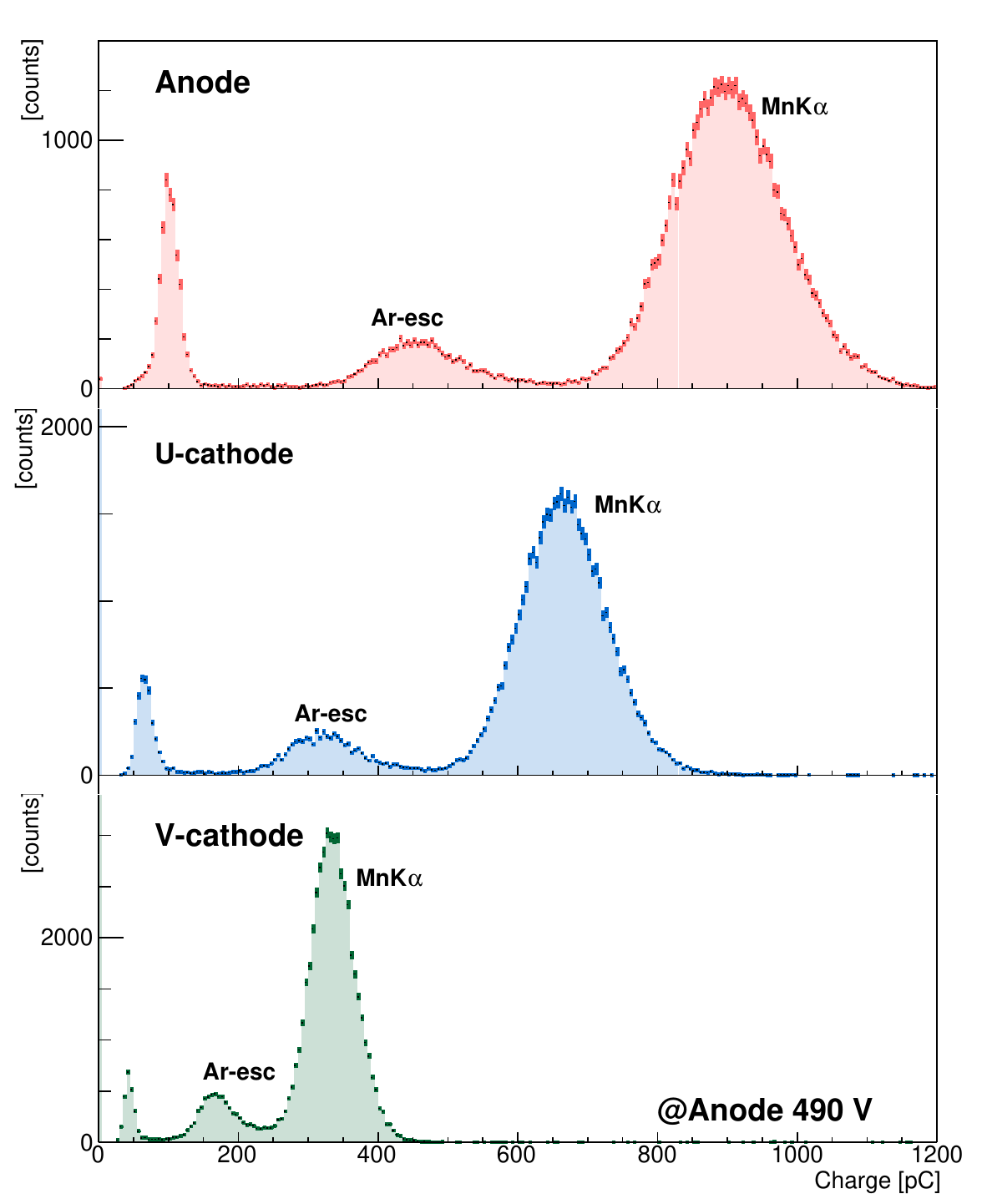}
    \caption{
        Charge distributions of the anode (top panel), U-cathode (middle panel) 
        and V-cathode (bottom panel), respectively, 
        using the $^{55}$Fe source at the anode voltage of 490~V.
        The most left-side peaks of each panel represent photoabsorption events of 5.9~keV X-ray 
        between the GEM and the triaxial \uPIC.
    }
    \label{fig:spectrum}
\end{figure}

\begin{figure}
    \centering
    \includegraphics[width=\linewidth]{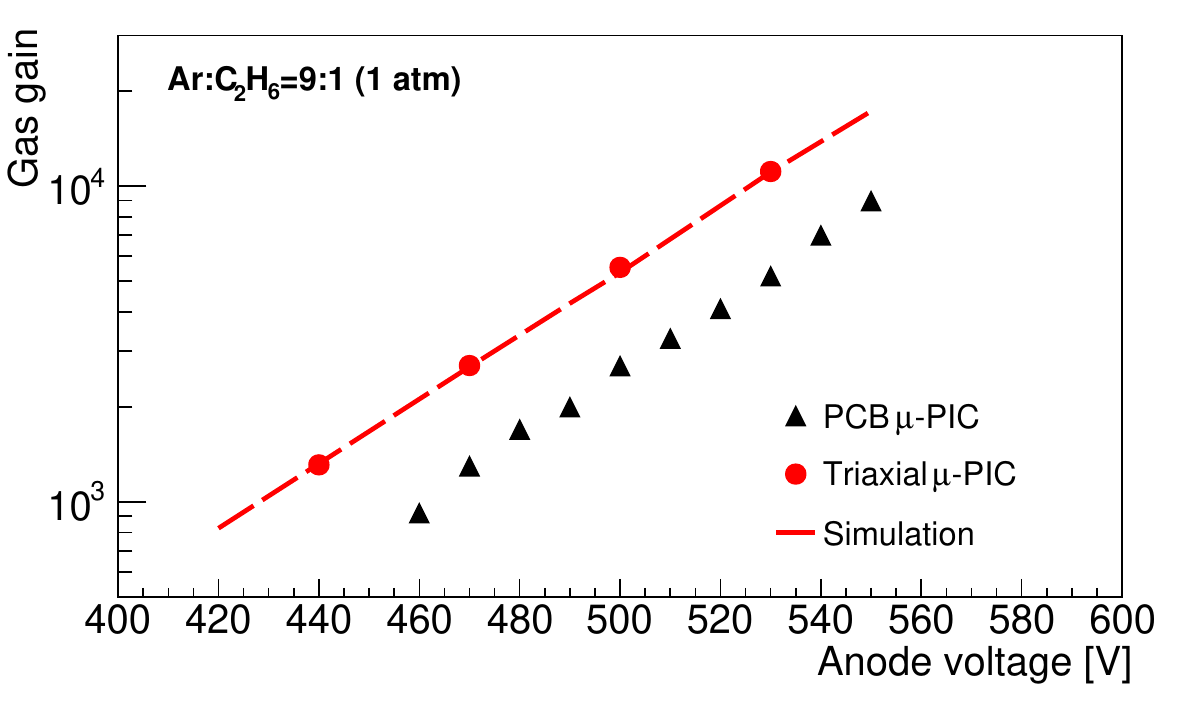}
    \caption{
        Gas gain of the triaxial \uPIC as a function of anode voltage. 
        The circles and triangles are the measurements of the triaxial \uPIC and the PCB \uPIC, respectively.
        The simulation result of the triaxial \uPIC by the Garfield++ simulation is also shown 
        as the dashed red line.
        Note that these values represent the net gas gain of the \uPIC itself, excluding the gain contributed by the GEM.
    }
    \label{fig:gain_curve_anode}
\end{figure}

Figure~\ref{fig:waveform} shows the waveforms obtained 
at an anode voltage of 490~V using a digital oscilloscope
under irradiating X-rays from $^{55}$Fe.
Reverse-polarity signals from the V-cathode (blue) and U-cathode (magenta) were 
observed simultaneously with the anode signal (yellow). 
The pulse heights of the two cathodes are comparable, 
and a shorter decay time was observed for the V-cathode compared to the anode and U-cathode signals, 
which is consistent with the Garfield++ simulations.
Figure~\ref{fig:spectrum} shows the charge distribution spectra 
of the anode, U-cathode, and V-cathode at an anode voltage of 490~V. 
The 5.9~keV photopeak and the 2.9~keV Ar escape peak were clearly observed for all electrodes. 
The measured gas gain of the anode in the triaxial \uPIC was $3.9\times10^3$. 
The energy resolutions (FWHM) for the anode, V-cathode, and U-cathode were 
21\%, 23\%, and 22\%, respectively. 
The charge ratio was found to be 3:2:1 (X:U:V), which is consistent with the simulated design. 
These results confirm that the triaxial \uPIC operates as expected.
The measured gas gain as a function of the anode voltage is shown in Figure~\ref{fig:gain_curve_anode}. 
The gain of the triaxial \uPIC (circles) is in excellent agreement 
with the estimation from the Garfield++ simulation. 
Furthermore, it is higher than the gain of the PCB \uPIC (triangles), 
thanks to the 200-$\mu$m-thick insulating substrate. 
The operation remained highly stable throughout all measurements, 
with a discharge rate of less than 0.002~Hz.

\subsection{Resolving the multi-hit ambiguity}
\begin{figure}
    \centering
    \includegraphics[width=\linewidth]{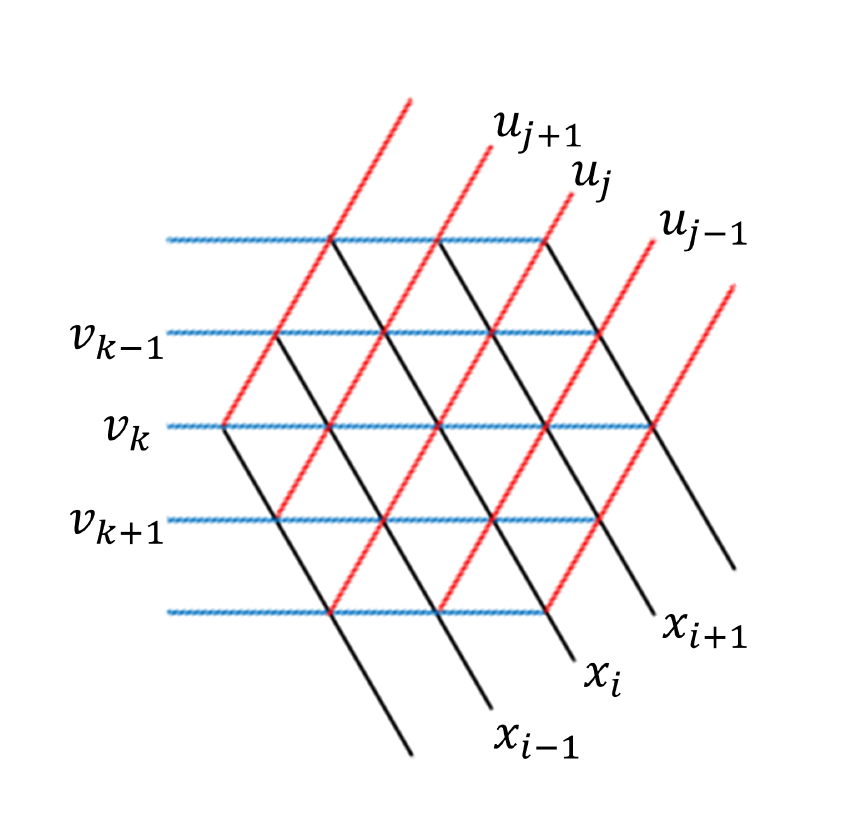}
    \caption{
        Definition of the identification number for the strip electrode: 
        When the coordinates of each axis are arranged at relative angular intervals of 120$^\circ$,
        the sum of the strip IDs from all three axes is invariant across all pixels.
    }
    \label{fig:axes_def}
\end{figure}

To determine the hit position, it is necessary to identify the coincidence of the X, U, and V-axis signals.
Naively, the calculation cost is $O(N^3)$, whereas that for an orthogonal \uPIC is $O(N^2)$,
where $N$ is the number of readout strip electrode.
However, when the coordinates of each axis are arranged in an axially symmetric configuration 
with a relative angular interval of 120$^\circ$, as shown in Figure~\ref{fig:axes_def},
the sum of the strip IDs from all three axes remains constant across all pixels:
\begin{equation}
    A_{ijk} \equiv i+j+k = const.
\end{equation}
The triaxial \uPIC fabricated in this study has a constant sum $A_{ijk}=192$, 
where the strip IDs for the X, U, and V axes range from 0 to 127.
To reduce the computational load, 
we determined the hit positions by exploiting this characteristic of the triaxial \uPIC.
First, we identify the coincidence between the U-cathode and V-cathode strips.
If the coincident pixel represents a true hit, 
the corresponding anode strip with the ID $A_{ijk}-(j+k)$ must also contain a signal. 
Therefore, we can distinguish whether a coincident pixel is a true hit or a fake hit 
by verifying the existence of a signal on that specific and unique anode strip.
Using this method (3-axis method), the computational cost is reduced to $O(N^2)$. 
The coordinates of the hit pixel $(x, y)$ are determined as follows:
\begin{align}
    x &= 0.48 \left( i-\frac{\max(i)}{2} \right) \quad \text{[mm]}, \\
    y &= \frac{0.48}{\sqrt{3}} \left( j - k \right) \quad \text{[mm]}.    
\end{align}
where $i$, $j$ and $k$ are the strip IDs of the corresponding axes.

\begin{figure}
    \centering
    \includegraphics[width=\linewidth]{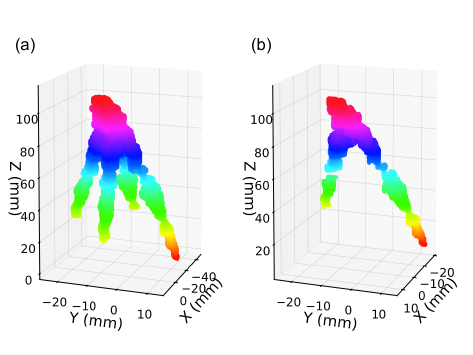}
    \caption{
        Reconstructed track image of the pair creation by the 2-axis coincidence with U and V-cathode (a) and 3-axis method (b).
        The color gradation indicates the height of $z$-axis.
    }
    \label{fig:pair_track}
\end{figure}

A portion of the full-absorption events from $^{55}$Fe generates two separate clusters 
due to the photoabsorption electron and the subsequent de-excitation X-ray. 
Since these clusters arrive simultaneously within a 1~cm drift length, 
track reconstruction fails when using a conventional time-coincidence method; 
specifically, four clouds are reconstructed in a dual-axis \uPIC (such as PCB or TGV types). 
Thus, these two-cluster events from $^{55}$Fe effectively represent the multi-hit ambiguity.
\begin{figure}
    \centering
    \includegraphics[width=\linewidth]{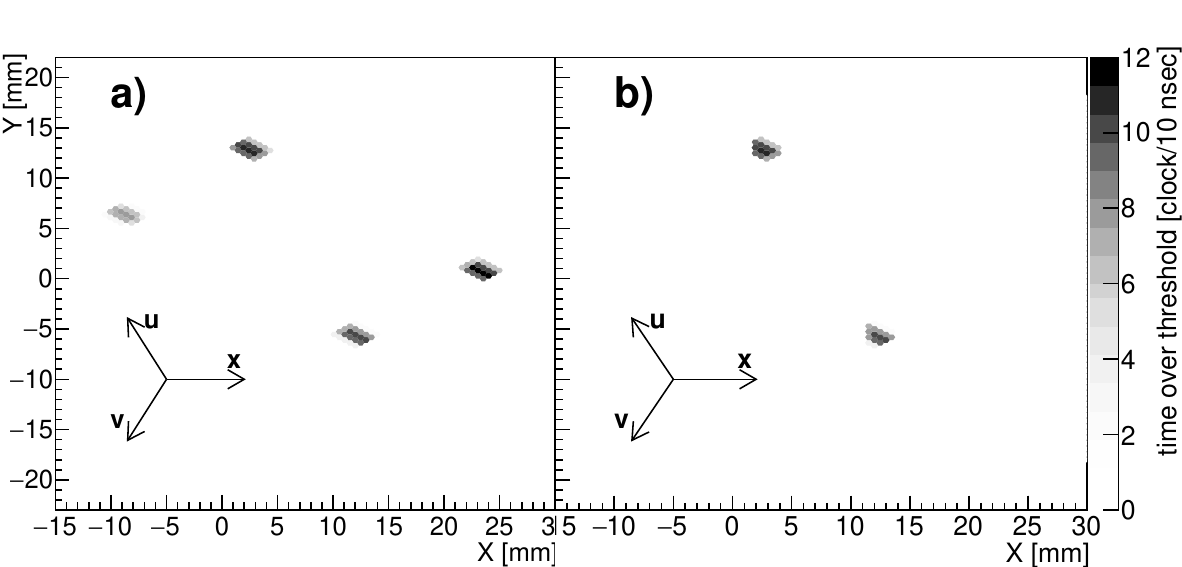}
    \caption{
        (a) Two cluster reconstruction by the 2-axis coincidence with U and V-cathode. 
        Two cluster out of four are fake tracks.
        (b) Two cluster reconstruction by the 3-axis coincidence.
        The fake tracks are eliminated.
    }
    \label{fig:cluster_image}
\end{figure}
We demonstrated two-cluster reconstruction using 3-axis method
involving the anode, V-cathode, and U-cathode. 
Figure~\ref{fig:cluster_image} shows the resulting image compared 
with the reconstruction using the 2-axis method. 
We successfully eliminated the two ghost clouds thanks to the additional axis information.
Furthermore, we evaluated the resolving accuracy. 
In this study, a track consisting of adjacent pixel hits is defined as a cluster. 
Since most 3.0~keV clusters have a radius of 1.5~mm, 
clusters separated by more than 3~mm are recognized as distinct. 
From the full-absorption events, 
384 instances of four-cloud reconstructions were observed using the 2-axis coincidence method. 
When the 3-axis method was applied, 
all of these multi-cluster events were correctly resolved as two-cluster events. 
Consequently, the probability of a false hit is less than 0.78\% at a 95\% confidence level
for two clusters separated by a distance of over 3 mm.
Finally, Figure~\ref{fig:pair_track} presents a three-dimensional image of a pair creation event 
captured using a time projection chamber with a 10~cm drift length. 
Similar to the $^{55}$Fe results, ghost tracks were successfully eliminated using the 3-axis method,
resolving the multi-hit ambiguity. 
These results demonstrate that the triaxial \uPIC is a powerful tool 
for accurately reconstructing multi-hit track events.

\section{High gain operation}
\label{sec:alt_op}
In the fundamental tests described in the previous section, 
the triaxial \uPIC was operated with both the U- and V-cathodes connected to the ground via a 10~M$\Omega$ resistor. 
In this configuration, 
the U-cathode functioned primarily as a readout electrode to resolve multi-hit ambiguity. 
However, the triaxial \uPIC is also expected to generate a stronger electric field 
near the anode by applying a negative potential to the U-cathode. 
In this section, we discuss the high-gain operation achieved by biasing the U-cathode.

\begin{figure}
    \centering
    \includegraphics[width=\linewidth]{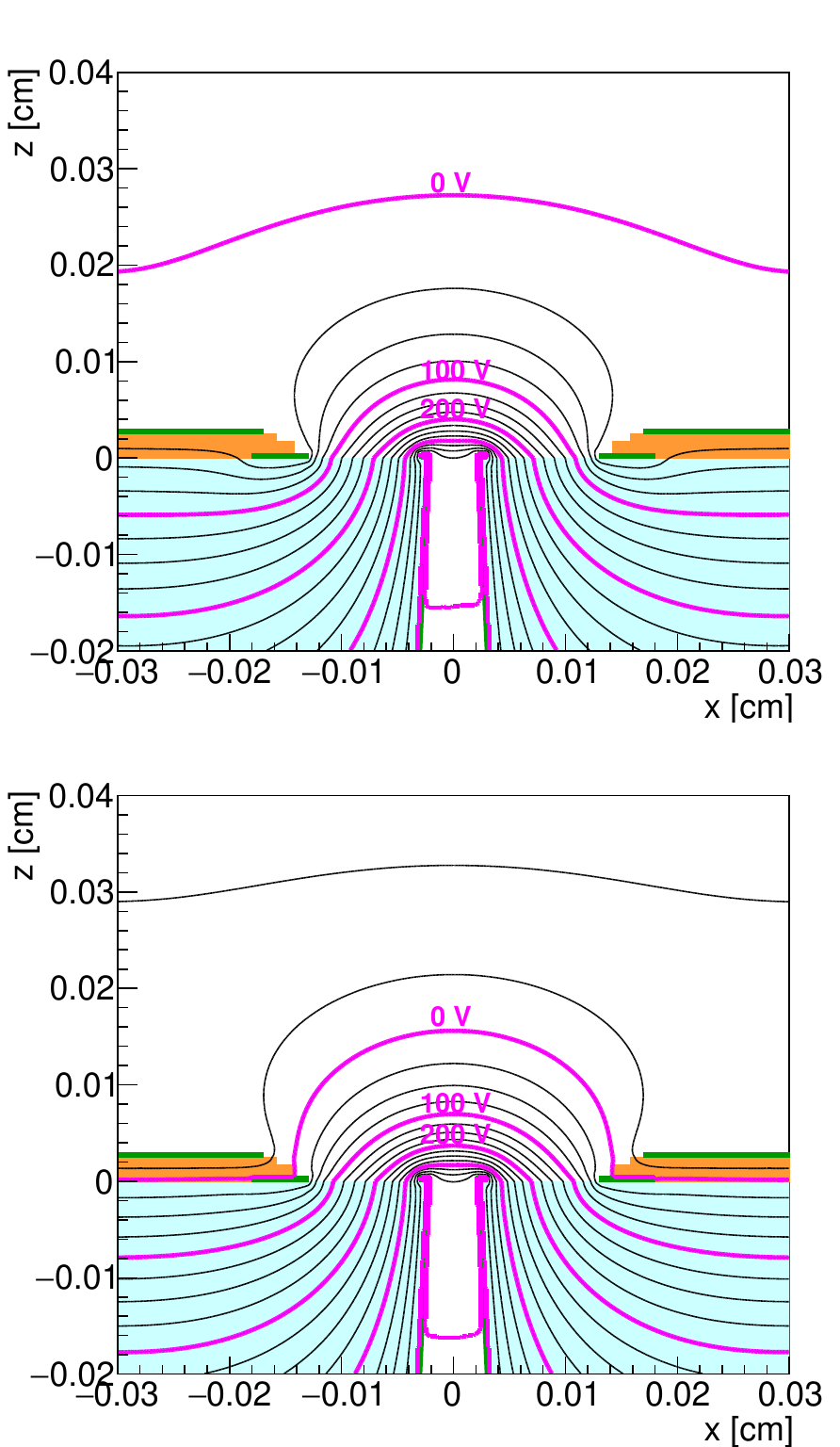}
    \caption{
        Simulated equipotential lines with the U-cathode Voltage of 0~V (Top) and -50~V (Bottom).
        The potentials of the anode and V-cathode were fixed at 400~V and 0~V, respectively.
    }
    \label{fig:electric_field}
\end{figure}

\begin{figure}
    \centering
    \includegraphics[width=\linewidth]{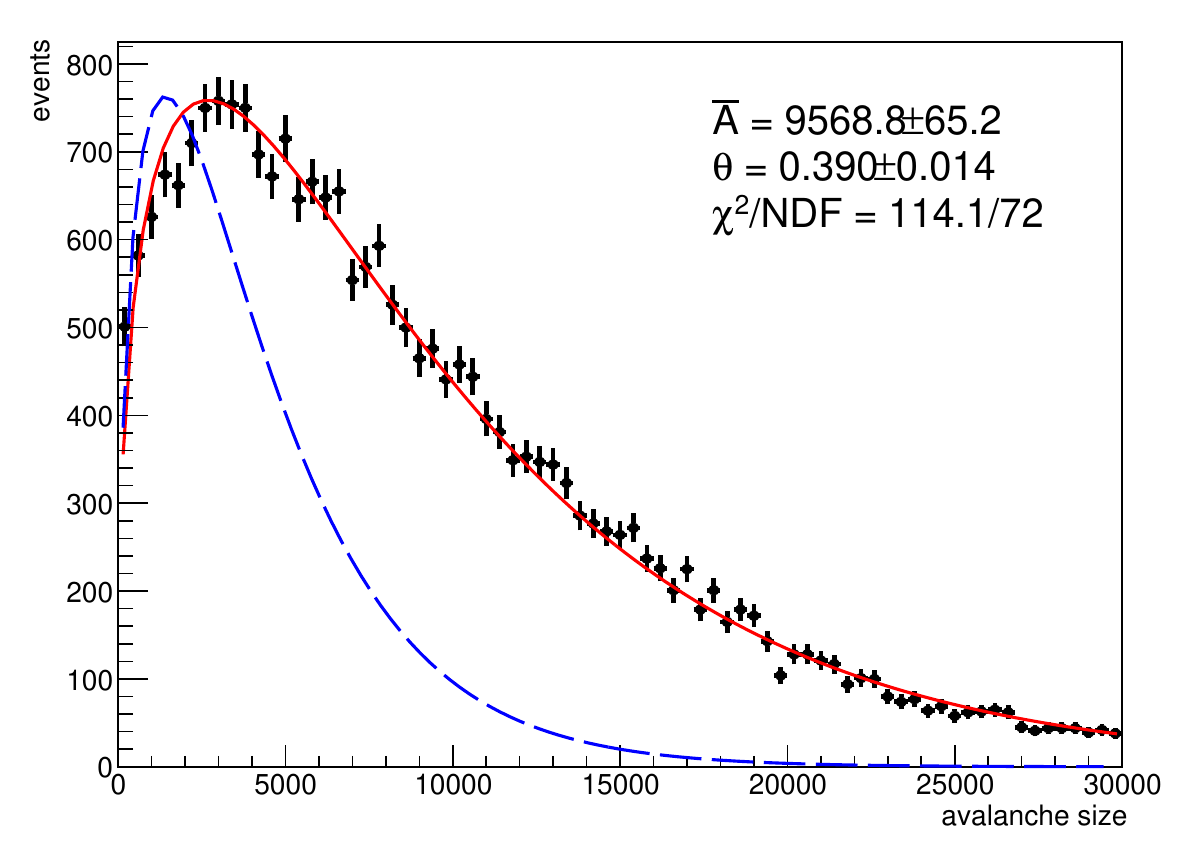}
    \caption{
        Simulated single-electron spectrum of a triaxial \uPIC 
        with an anode, U- and V-cathode voltage of 490~V -50~V and 0~V, respectively.
        Assumed gas is Ar 90\% + C$_2$H$_6$ 10\% at a pressure of 1~atm.
        The solid line is the best fit with a Polya distribution.
        The dashed line is a single-electron spectrum shown in Figure~\ref{fig:avalanche_single}.
    }
    \label{fig:as_Um50}
\end{figure}

\begin{figure}
    \centering
    \includegraphics[width=\linewidth]{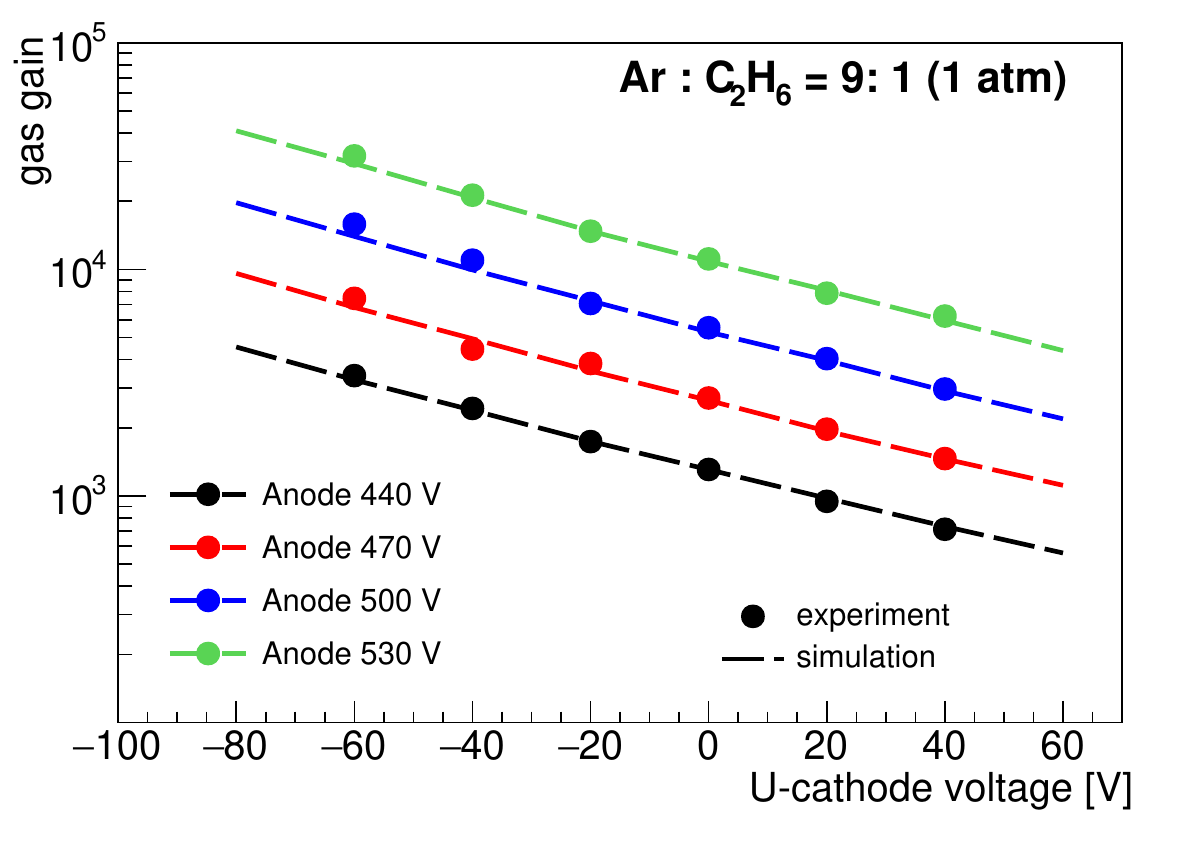}
    \caption{
        Net gas gain of the \uPIC (excluding the GEM gain) as a function of the electric potential applied to the U-cathode. 
        The black, red, blue and green points indicates 
        the anode of 440 V, 470 V, 500 V and 530 V, respectively. 
        The dashed lines are the simulation results from Garfield++
    }
    \label{fig:gaincurve_xu}
\end{figure}

We simulated the electric field in the vicinity of the anode using Elmer,
comparing cases where potentials of 0~V and -50~V were applied to the U-cathode.
The potentials of the anode and V-cathode were fixed at 400~V and 0~V, respectively.
The resulting two-dimensional equipotential line distributions are shown in Figure~\ref{fig:electric_field}.
With a U-cathode potential of -50~V, 
the equipotential lines are more concentrated near the anode compared to the 0~V case.
Figure~\ref{fig:as_Um50} shows the simulated single-electron spectrum with a U-cathode voltage of -50~V.
Compared to the basic operation at 0~V, the simulated gas gain is 2.3 times higher. 
These simulation results indicate that applying a negative potential 
to the U-cathode enables more effective gas amplification.
To confirm the high-gain operation, 
we measured the gas gain at various U-cathode voltages using the configuration
shown in Figure~\ref{fig:schematic}, under irradiation of an $^{55}$Fe X-ray source. 
The obtained gas gain as a function of the U-cathode voltage is presented in Figure~\ref{fig:gaincurve_xu}.
As expected, we confirmed that the negative electric potential of the U-cathode provided additional gas gain,
which is consistent with the Garfield++ simulation results. 
The charge ratio among the three electrodes (X:U:V = 3:2:1) remained almost constant across all measurements,
consistent with the basic operation. 
The operation was stable up to a maximum gas gain of $3.2\times 10^4$.
These results demonstrate 
that a slight increase in the U-cathode voltage significantly enhances the gas gain by several times, 
while maintaining a low discharge rate.

\section{Summary}
Various \uPIC applications require resolving multi-hit ambiguity 
to trace the winding tracks of charged particles or for high-rate operation. 
For this purpose, 
we introduced an additional readout axis and designed a new electrode structure 
featuring vertically divided concentric cathodes. 
The anode, U-cathode, and V-cathode are connected to their respective readout strip electrodes, 
with differential angles of 120$^\circ$ between the strips.
Garfield++ simulations indicated that the readout charge ratio reaches X:U:V = 3:2:1, 
and the pulse heights of the U- and V-cathodes are almost identical 
when using a CSA with a time constant of approximately 100~ns.
In this study, we fabricated a triaxial \uPIC based on the TGV \uPIC, 
featuring a hexagonal shape with a diagonal length of approximately 7~cm.
Under irradiation from an $^{55}$Fe X-ray source, 
we demonstrated that the output signals of this triaxial \uPIC are 
in good agreement with the Garfield++ simulations. 
The obtained gas gain reached $1.1\times 10^4$ at an anode voltage of 530~V,
with an energy resolution of 22\% (FWHM) for 5.9~keV.
To confirm the resolution of multi-hit ambiguity, 
we analyzed multi-track events and found the probability of a false hit 
to be less than 0.78\% at a 95\% confidence level.
Furthermore, applying a negative voltage to the U-cathode is expected 
to yield a higher gas gain than in nominal operation. 
We demonstrated this high-gain operation through measurements under $^{55}$Fe X-ray irradiation, 
and the triaxial \uPIC performed as predicted by the Garfield++ simulations.
Without increasing the risk of discharge,
the gas gain reached $3.2\times 10^4$ with the anode, U-, and V-cathode voltages 
set to 530~V, -60~V, and 0~V, respectively.
The size of the triaxial \uPIC fabricated in this study was limited by the dimensions of the glass wafer. 
If larger glass wafers can be produced, 
the triaxial \uPIC will be effective in various TPC applications.

\section*{Acknowledgments}
We would like to express our gratitude to the late A. Ochi 
for his contributions and insights into the development of the \uPIC.
This study was supported by JSPS KAKENHI Grant Number JP25K24556.

\bibliography{my_bib}

@ARTICLE{2001NIMPA.471..264O,
       author = {{Ochi}, Atsuhiko and {Nagayoshi}, Tsutomu and {Koishi}, Satoshi and {Tanimori}, Toru and {Nagae}, Tomofumi and {Nakamura}, Mirei},
        title = "{A new design of the gaseous imaging detector: Micro Pixel Chamber}",
      journal = {Nuclear Instruments and Methods in Physics Research A},
         year = 2001,
        month = sep,
       volume = {471},
       number = {1-2},
        pages = {264-267},
          doi = {10.1016/S0168-9002(01)00996-2}
}

@ARTICLE{2004NIMPA.535..236M,
       author = {{Miuchi}, Kentaro and {Kubo}, Hidetoshi and {Nagayoshi}, Tsutomu and {Okada}, Yoko and {Orito}, Reiko and {Takada}, Atsushi and {Takeda}, Atsushi and {Tanimori}, Toru and {Ueno}, Masaru and {Bouianov}, Oleg and {Bouianov}, Marina},
        title = "{Performance and applications of a {\ensuremath{\mu}}-TPC}",
      journal = {Nuclear Instruments and Methods in Physics Research A},
         year = 2004,
        month = dec,
       volume = {535},
        pages = {236-241},
          doi = {10.1016/j.nima.2004.07.143}
}

@ARTICLE{2004NIMPA.525...20N,
       author = {{Nagayoshi}, Tsutomu and {Kubo}, Hidetoshi and {Miuchi}, Kentaro and {Orito}, Reiko and {Takada}, Atsushi and {Takeda}, Atsushi and {Tanimori}, Toru and {Ueno}, Masaru and {Bouianov}, Oleg and {Bouianov}, Marina},
        title = "{Development of {\ensuremath{\mu}}-PIC and its imaging properties}",
      journal = {Nuclear Instruments and Methods in Physics Research A},
         year = 2004,
        month = jun,
       volume = {525},
       number = {1-2},
        pages = {20-27},
          doi = {10.1016/j.nima.2004.03.019}
}

@ARTICLE{2007NIMPA.573..195T,
       author = {{Takada}, A. and {Hattori}, K. and {Kabuki}, S. and {Kubo}, H. and {Miuchi}, K. and {Nagayoshi}, T. and {Nishimura}, H. and {Okada}, Y. and {Orito}, R. and {Sekiya}, H. and {Takeda}, A. and {Tanimori}, T. and {Ueno}, K.},
        title = "{A very large area Micro Pixel Chamber}",
      journal = {Nuclear Instruments and Methods in Physics Research A},
         year = 2007,
        month = apr,
       volume = {573},
       number = {1-2},
        pages = {195-199},
          doi = {10.1016/j.nima.2006.10.283}
}

@ARTICLE{2004NewAR..48..263T,
      author = {{Tanimori}, T. and {Kubo}, H. and {Miuchi}, K. and {Nagayoshi}, T. and {Orito}, R. and {Takada}, A. and {Takeda}, A. and {Ueno}, M.},
        title = "{MeV {\ensuremath{\gamma}}-ray imaging detector with micro-TPC}",
      journal = {New Astronomy Reviews},
         year = 2004,
        month = feb,
      volume = {48},
      number = {1-4},
        pages = {263-268},
          doi = {10.1016/j.newar.2003.11.038}
}

@ARTICLE{2004NIMPA.529..373T,
      author = {{Tanimori}, Toru and {Kubo}, Hidetoshi and {Miuchi}, Kentaro and {Nagayoshi}, Tsutomu and {Okada}, Yoko and {Orito}, Reiko and {Takada}, Atsushi and {Takeda}, Atsushi},
        title = "{Time projection chamber based on micro-pattern detector for neutron time-resolved imaging}",
      journal = {Nuclear Instruments and Methods in Physics Research A},
         year = 2004,
        month = aug,
      volume = {529},
      number = {1-3},
        pages = {373-377},
          doi = {10.1016/j.nima.2004.05.014}
}

@ARTICLE{2007PhLB..654...58M,
       author = {{Miuchi}, Kentaro and {Hattori}, Kaori and {Kabuki}, Shigeto and {Kubo}, Hidetoshi and {Kurosawa}, Shunsuke and {Nishimura}, Hironobu and {Okada}, Yoko and {Takada}, Atsushi and {Tanimori}, Toru and {Tsuchiya}, Ken'ichi and {Ueno}, Kazuki and {Sekiya}, Hiroyuki and {Takeda}, Atsushi},
        title = "{Direction-sensitive dark matter search results in a surface laboratory}",
      journal = {Physics Letters B},
         year = 2007,
        month = oct,
       volume = {654},
       number = {3-4},
        pages = {58-64},
          doi = {10.1016/j.physletb.2007.08.042}
}

@ARTICLE{2010NIMPA.623..606K,
       author = {{Kabuki}, Shigeto and {Kimura}, Hiroyuki and {Amano}, Hiroo and {Nakamoto}, Yuji and {Kubo}, Hidetoshi and {Miuchi}, Kentaro and {Kurosawa}, Shunsuke and {Takahashi}, Michiaki and {Kawashima}, Hidekazu and {Ueda}, Masashi and {Okada}, Tomohisa and {Kubo}, Atsushi and {Kunieda}, Etuso and {Nakahara}, Tadaki and {Kohara}, Ryota and {Miyazaki}, Osamu and {Nakazawa}, Tetsuo and {Shirahata}, Takashi and {Yamamoto}, Etsuji and {Ogawa}, Koichi and {Togashi}, Kaori and {Saji}, Hideo and {Tanimori}, Toru},
        title = "{Electron-tracking Compton gamma-ray camera for small animal and phantom imaging}",
      journal = {Nuclear Instruments and Methods in Physics Research A},
         year = 2010,
        month = nov,
       volume = {623},
       number = {1},
        pages = {606-607},
          doi = {10.1016/j.nima.2010.03.085}
}

@ARTICLE{2013NIMPA.732..591K,
       author = {{Kishimoto}, Yuji and {Sasaki}, Shinichi and {Saito}, Kiwamu and {Takahashi}, Kazutoshi and {Doke}, Tadayoshi and {Miuchi}, Kentaro and {Fuse}, Tetsuhito and {Nagamatsu}, Aiko and {Uchihori}, Yukio and {Kitamura}, Hisashi and {Terasawa}, Kazuhiro},
        title = "{Basic performance of a position-sensitive tissue-equivalent proportional chamber (PS-TEPC)}",
      journal = {Nuclear Instruments and Methods in Physics Research A},
         year = 2013,
        month = dec,
       volume = {732},
        pages = {591-594},
          doi = {10.1016/j.nima.2013.05.149}
}

@INPROCEEDINGS{2014JPhCS.569a2042F,
       author = {{Furuno}, T. and {Kawabata}, T. and {Adachi}, S. and {Ayyad}, Y. and {Baba}, T. and {Hashimoto}, T. and {Ishii}, Y. and {Kabuki}, S. and {Kubo}, H. and {Matsuda}, Y. and {Matsuoka}, Y. and {Mizumoto}, T. and {Murata}, M. and {Ong}, H.~J. and {Sawano}, T. and {Takada}, A. and {Tanaka}, J. and {Tanihata}, I. and {Tanimori}, T. and {Tsumura}, M. and {Watanabe}, H.~D.},
        title = "{Development of MAIKo: the active target with {\ensuremath{\mu}}-PIC for RI beam experiments}",
    booktitle = {Journal of Physics Conference Series},
         year = 2014,
       series = {Journal of Physics Conference Series},
       volume = {569},
        month = dec,
          eid = {012042},
        pages = {012042},
          doi = {10.1088/1742-6596/569/1/012042}
}

@ARTICLE{2017NatSR...741972T,
       author = {{Tomono}, Dai and {Mizumoto}, Tetsuya and {Takada}, Atsushi and {Komura}, Shotaro and {Matsuoka}, Yoshihiro and {Mizumura}, Yoshitaka and {Oda}, Makoto and {Tanimori}, Toru},
        title = "{First On-Site True Gamma-Ray Imaging-Spectroscopy of Contamination near Fukushima Plant}",
      journal = {Scientific Reports},
         year = 2017,
        month = feb,
       volume = {7},
          eid = {41972},
        pages = {41972},
          doi = {10.1038/srep41972}
}

@ARTICLE{2005NIMPA.546..457N,
       author = {{Nagayoshi}, Tsutomu and {Takada}, Atsushi and {Kubo}, Hidetoshi and {Miuchi}, Kentaro and {Orito}, Reiko and {Okada}, Yoko and {Takeda}, Atsushi and {Tanimori}, Toru and {Ueno}, Masaru and {Bouianov}, Oleg and {Bouianov}, Marina},
        title = "{Simulation study of electron drift and gas multiplication in Micro Pixel Chamber}",
      journal = {Nuclear Instruments and Methods in Physics Research A},
         year = 2005,
        month = jul,
       volume = {546},
       number = {3},
        pages = {457-465},
          doi = {10.1016/j.nima.2005.02.042}
}

@INPROCEEDINGS{2018EPJWC.17402010T,
       author = {{Takemura}, T. and {Takada}, A. and {Kishimoto}, T. and {Komura}, S. and {Kubo}, H. and {Matsuoka}, Y. and {Miuchi}, K. and {Miyamoto}, S. and {Mizumoto}, T. and {Mizumura}, Y. and {Motomura}, T. and {Nakamasu}, Y. and {Nakamura}, K. and {Oda}, M. and {Ohta}, K. and {Parker}, J.~D. and {Sawano}, T. and {Sonoda}, S. and {Tanimori}, T. and {Tomono}, D. and {Yoshikawa}, K.},
        title = "{Development of the micro pixel chamber based on MEMS technology}",
    booktitle = {European Physical Journal Web of Conferences},
         year = 2018,
       series = {European Physical Journal Web of Conferences},
       volume = {174},
        month = feb,
          eid = {02010},
        pages = {02010},
          doi = {10.1051/epjconf/201817402010}
}

@INPROCEEDINGS{2020JPhCS1498a2002A,
       author = {{Abe}, M. and {Tanimori}, T. and {Takada}, A. and {Mizumura}, Y. and {Komura}, S. and {Kishimoto}, T. and {Takemura}, T. and {Yoshikawa}, K. and {Nakamura}, Y. and {Nakamasu}, Y. and {Taniguchi}, T. and {Onozaka}, K. and {Saito}, K. and {Mizumoto}, T. and {Sonoda}, S. and {Parker}, J.~D. and {Miuchi}, K. and {Sawano}, T.},
        title = "{Development of a {\ensuremath{\mu}}-PIC with glass substrate aiming at high gas gain}",
    booktitle = {Journal of Physics Conference Series},
         year = 2020,
       series = {Journal of Physics Conference Series},
       volume = {1498},
        month = apr,
          eid = {012002},
        pages = {012002},
          doi = {10.1088/1742-6596/1498/1/012002}
}

@ARTICLE{2005NIMPA.540..266B,
       author = {{Bouianov}, Marina and {Bouianov}, Oleg and {Nagayoshi}, Tsutomu and {Kubo}, Hidetoshi and {Miuchi}, Kentaro and {Orito}, Reiko and {Takada}, Atsushi and {Takeda}, Atsushi and {Tanimori}, Toru},
        title = "{Performance optimisation of the micro pixel chamber}",
      journal = {Nuclear Instruments and Methods in Physics Research A},
         year = 2005,
        month = mar,
       volume = {540},
       number = {2-3},
        pages = {266-272},
          doi = {10.1016/j.nima.2004.11.041}
}

@ARTICLE{2013JInst...8C0023T,
       author = {{Takada}, A. and {Tanimori}, T. and {Kubo}, H. and {Parker}, J.~D. and {Mizumoto}, T. and {Mizumura}, Y. and {Iwaki}, S. and {Sawano}, T. and {Nakamura}, K. and {Taniue}, K. and {Higashi}, N. and {Matsuoka}, Y. and {Komura}, S. and {Sato}, Y. and {Namamura}, S. and {Oda}, M. and {Sonoda}, S. and {Tomono}, D. and {Miuchi}, K. and {Kabuki}, S. and {Kishimoto}, Y. and {Kurosawa}, S.},
        title = "{Simulation of gas avalanche in a micro pixel chamber using Garfield++}",
      journal = {Journal of Instrumentation},
         year = 2013,
        month = oct,
       volume = {8},
       number = {10},
          eid = {C10023},
        pages = {C10023},
          doi = {10.1088/1748-0221/8/10/C10023}
}

@ARTICLE{1997NIMPA.386..531S,
       author = {{Sauli}, F.},
        title = "{GEM: A new concept for electron amplification in gas detectors}",
      journal = {Nuclear Instruments and Methods in Physics Research A},
         year = 1997,
        month = feb,
       volume = {386},
       number = {2},
        pages = {531-534},
          doi = {10.1016/S0168-9002(96)01172-2}
}

@ARTICLE{2009NIMPA.608..390T,
       author = {{Tamagawa}, T. and {Hayato}, A. and {Asami}, F. and {Abe}, K. and {Iwamoto}, S. and {Nakamura}, S. and {Harayama}, A. and {Iwahashi}, T. and {Konami}, S. and {Hamagaki}, H. and {Yamaguchi}, Y.~L. and {Tawara}, H. and {Makishima}, K.},
        title = "{Development of thick-foil and fine-pitch GEMs with a laser etching technique}",
      journal = {Nuclear Instruments and Methods in Physics Research A},
         year = 2009,
        month = sep,
       volume = {608},
       number = {3},
        pages = {390-396},
          doi = {10.1016/j.nima.2009.07.014}
}

@ARTICLE{2015NIMPA.800...40M,
       author = {{Mizumoto}, T. and {Matsuoka}, Y. and {Mizumura}, Y. and {Tanimori}, T. and {Kubo}, H. and {Takada}, A. and {Iwaki}, S. and {Sawano}, T. and {Nakamura}, K. and {Komura}, S. and {Nakamura}, S. and {Kishimoto}, T. and {Oda}, M. and {Miyamoto}, S. and {Takemura}, T. and {Parker}, J.~D. and {Tomono}, D. and {Sonoda}, S. and {Miuchi}, K. and {Kurosawa}, S.},
        title = "{New readout and data-acquisition system in an electron-tracking Compton camera for MeV gamma-ray astronomy (SMILE-II)}",
      journal = {Nuclear Instruments and Methods in Physics Research A},
         year = 2015,
        month = nov,
       volume = {800},
        pages = {40-50},
          doi = {10.1016/j.nima.2015.08.004}
}

@ARTICLE{2002NIMPA.478..104B,
       author = {{Bachmann}, S. and {Kappler}, S. and {Ketzer}, B. and {M{\"u}ller}, Th. and {Ropelewski}, L. and {Sauli}, F. and {Schulte}, E.},
        title = "{High rate X-ray imaging using multi-GEM detectors with a novel readout design}",
      journal = {Nuclear Instruments and Methods in Physics Research A},
         year = 2002,
        month = feb,
       volume = {478},
       number = {1-2},
        pages = {104-108},
          doi = {10.1016/S0168-9002(01)01719-3}
}

@ARTICLE{2005NIMPA.553...18S,
       author = {{Sauli}, Fabio},
        title = "{Novel Cherenkov photon detectors}",
      journal = {Nuclear Instruments and Methods in Physics Research A},
         year = 2005,
        month = nov,
       volume = {553},
       number = {1-2},
        pages = {18-24},
          doi = {10.1016/j.nima.2005.08.024}
}

@ARTICLE{1998NIMPA.419..726V,
       author = {{Veenhof}, Rob},
        title = "{Garfield, recent developments}",
      journal = {Nuclear Instruments and Methods in Physics Research A},
         year = 1998,
        month = dec,
       volume = {419},
       number = {2-3},
        pages = {726-730},
          doi = {10.1016/S0168-9002(98)00851-1}
}

@ARTICLE{Garfield_pp,
        title = "{Garfield++}",
          url = {https://garfieldpp.web.cern.ch/}
}

@ARTICLE{Gmsh_HP,
        title = "{Gmsh}",
          url = {https://gmsh.info/}
}

@ARTICLE{Elmer_HP,
        title = "{Elmer}",
          url = {https://www.elmerfem.org/blog/}
}

@ARTICLE{1996NIMPA.381..280T,
       author = {{Tanimori}, Toru and {Ochi}, Atsuhiko and {Minami}, Seiji and {Nagae}, Tomofumi},
        title = "{Development of an imaging microstrip gas chamber with a 5 cm {\texttimes} 5 cm area based on multi-chip module technology}",
      journal = {Nuclear Instruments and Methods in Physics Research A},
         year = 1996,
        month = feb,
       volume = {381},
       number = {2},
        pages = {280-288},
          doi = {10.1016/S0168-9002(96)00833-9}
}

@ARTICLE{2022ApJ...930....6T,
       author = {{Takada}, Atsushi and {Takemura}, Taito and {Yoshikawa}, Kei and {Mizumura}, Yoshitaka and {Ikeda}, Tomonori and {Nakamura}, Yuta and {Onozaka}, Ken and {Abe}, Mitsuru and {Hamaguchi}, Kenji and {Kubo}, Hidetoshi and {Kurosawa}, Shunsuke and {Miuchi}, Kentaro and {Saito}, Kaname and {Sawano}, Tatsuya and {Tanimori}, Toru},
        title = "{First Observation of the MeV Gamma-Ray Universe with Bijective Imaging Spectroscopy Using the Electron-tracking Compton Telescope on Board SMILE-2+}",
      journal = {The Astrophysucal Journal},
         year = 2022,
        month = may,
       volume = {930},
       number = {1},
          eid = {6},
        pages = {6},
          doi = {10.3847/1538-4357/ac6103},
archivePrefix = {arXiv},
       eprint = {2107.00180},
 primaryClass = {astro-ph.HE},
       adsurl = {https://ui.adsabs.harvard.edu/abs/2022ApJ...930....6T}
}

\end{document}